\documentclass[twocolumn,showpacs,preprintnumbers,amsmath,amssymb]{revtex4}

\usepackage{graphicx}
\usepackage{dcolumn}
\usepackage{bm}

\begin{document}

\title{Ignition and Propagation of Magnetic Avalanches in Mn$_{12}$-Acetate: the effect of quantum tunneling}

\author{S. McHugh$^1$}
\author{R. Jaafar$^1$}
\author{M.P. Sarachik$^1$}
\author{Y. Myasoedov$^2$}
\author{A. Finkler$^2$}
\author{H. Shtrikman$^2$}
\author{E. Zeldov$^2$}
\author{R. Bagai$^3$}
\author{G. Christou$^3$}
\affiliation{
$^1$Department of Physics, City College of New York, CUNY, New York, New York 10031, USA\\
$^2$Department of Condensed Matter Physics, The Weizmann Institute of Science, Rehovot 76100, Israel\\
$^3$Department of Chemistry, University of Florida, Gainesville, Florida 32611, USA}

\date{\today}

\begin{abstract}
Using a wire heater to ignite magnetic avalanches in fixed magnetic field applied along the easy axis of single crystals of the molecular magnet Mn$_{12}$-acetate, we report fast local measurements of the temperature and time-resolved measurements of the local magnetization as a function of magnetic field.   In addition to confirming maxima in the velocity of propagation, we find that avalanches trigger at a threshold temperature which exhibits pronounced minima at resonant magnetic fields, demonstrating that thermally assisted quantum tunneling plays an important role in the ignition as well as the propagation of magnetic avalanches in molecular magnets.
\end{abstract}


\maketitle

First synthesized in 1980  \cite{Lis}, Mn$_{12}$-acetate ([Mn$_{12}$O$_{12}$(CH$_3$COO)$_{16}$(H$_2$O)$_4$]$\cdot$2CH$_3$COOH$\cdot$4H$_2$O, hereafter referred to as Mn$_{12}$-ac), is a prototypical molecular magnet that is particularly interesting for its high spin and high bistable anisotropy  \cite{Sessoli}.  The magnetic core of each Mn$_{12}$-ac molecule is composed of 12 Mn atoms  strongly coupled to form a rigid spin $S=10$ cluster at low temperatures; strong uniaxial magnetic anisotropy along the tetragonal symmetry axis provides a $\approx 60$ K barrier against spin reversal and robust bistability at temperatures below the blocking temperature of $\approx 3$ K.  Composed of $\approx 10^{18}$ nominally identical magnetic molecules regularly arranged on a tetragonal lattice, Mn$_{12}$-ac samples have served as a platform for the study of a wide variety of interesting magnetic phenomena.   In particular, it was in this material that macroscopic quantum tunneling of the large spin $S=10$ was discovered  \cite{Friedman,Thomas} below the blocking temperature whenever a magnetic field applied parallel to the anisotropy axis brought into alignment a pair of energy levels on opposite sides of the anisotropy barrier corresponding to different spin projections, as illustrated in Fig. 1 (a)  (see Refs.
\onlinecite{Barbara, Gatteschi, Friedman2, del Barco} for reviews).

Although abrupt reversals of the magnetization, referred to as magnetic avalanches, have been regularly observed in molecular magnets \cite{Paulsen}, they received little attention until relatively recently.  Avalanches were thought to entail a thermal runaway process in which the reversing spins release heat, causing the relaxation of the remaining spins in the crystal  \cite{Paulsen}.  Indeed, both direct and indirect measurements of the heat released during an avalanche have confirmed their thermal nature \cite{thermal}.  Recent experiments of Suzuki et al.  \cite{Suzuki} have revealed that the magnetization reversal does not occur homogeneously throughout the sample, but travels instead with constant velocity as a narrow interface between regions of opposing magnetization.  In light of the thermal nature of the process and the relatively slow velocity of propagation of the interface ($\approx 10$ m/s), Suzuki {\it et al.}   \cite{Suzuki} have suggested this is "magnetic deflagration", in analogy with the very similar process of chemical combustion referred to as chemical defragration \cite{Glassman}.    

\begin{figure}[htbp]
\begin{center}
  \includegraphics[width=3.4in]{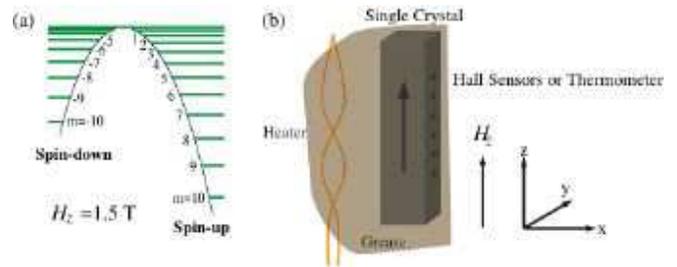}
\caption{(a) Energy level diagram for a longitudinal magnetic field of $1.5$ T.  (b) Schematic diagram of a crystal mounted on: (i) an array of Hall Sensors used to measure the magnetization, or (ii) a  germanium thermometer used to measure the temperature.  The heater, crystal, and sensors are all encased in Apiezon M grease.}
\label{default}
\end{center}
\end{figure}

Based on measurements of the time evolution of the total magnetization of  Mn$_{12}$-ac crystals during avalanches triggered by surface acoustic waves at fixed magnetic fields, Hernandez-Minguez et al.  \cite{Alberto2, Alberto3} have reported maxima in the velocity of propagation of avalanches in Mn$_{12}$-ac at "resonant" magnetic fields where the anisotropy barrier is effectively lowered by quantum tunneling of the spins.   The velocity maxima were attributed to thermally-assisted quantum deflagration.  

In the present paper we report the results of experiments designed to elucidate the role of quantum mechanics (i. e., spin-tunneling) in the ignition and propagation of magnetic avalanches in Mn$_{12}$-ac.  Using a wire heater to ignite magnetic avalanches in fixed magnetic field applied along the easy axis of single crystals of the molecular magnet Mn$_{12}$-acetate, we report fast local measurements of the temperature and time-resolved measurements of the local magnetization as a function of magnetic field.   We find that avalanches ignite at a reproducible threshold temperature, and this temperature exhibits pronounced minima at magnetic fields corresponding to thermally-assisted tunneling across the anisotropy barrier.  Additionally, we find maxima for the velocity of propagation of the avalanches, albeit in a higher range of magnetic field than reported by Hernandez-Minguez et al.  \cite{Alberto2, Alberto3}.

All measurements reported here were performed on single crystals of Mn$_{12}$-ac with typical dimensions of $1.5$ x $0.3$ x $0.3$ mm$^3$ immersed in liquid $^3$He at approximately $300$ mK.  Germanium thin film resistance thermometers of dimensions $40$ x $100  \mu$m$^2$ were deposited by e-gun evaporation on heated GaAs substrates in vacuum.  The crystal was mounted using a thin layer of thermally conductive Apiezon M grease (see Fig. 1b).  In order to make good thermal contact with the heater, the entire assembly, including thermometer, sample, and heater, was encased in Apiezon M grease, as shown in Fig. 1.  To minimize thermal gradients between the crystal and the thermometer, care was taken to place the heater as close as possible to the sample (roughly $1mm$ above the crystal) and the minimum heater power was used that still triggered avalanches.

It is well known that all crystals of Mn$_{12}$-ac contain a small amount of a second species of spin $S=10$ molecules that have a lower anisotropy barrier of roughly $45 K$.  This minor species is homogeneously distributed in the crystal at typical levels between $5$ and $8\%$   \cite{Sessoli, Wernsdorfer1}.  The presence of the minor species was found to have a significant influence on both the temperature for ignition and the propagation velocity of the avalanches.  Consequently, we used the following protocol to "quench" the effect of the minor species' spin relaxation  \cite{Wernsdorfer2}: after fully magnetizing the crystal in the "up" direction, the field was swept to a value in the opposite (downward) direction that is large enough to flip the minor species downward but small enough that it leaves the major species intact.  Bringing the magnetic field back to zero then yields a crystal with the major (spin "up") and minor (spin "down") species fully magnetized in opposite directions.  This allows the magnetic relaxation of minor and major species of Mn$_{12}$-ac to be studied independently.  We will report a detailed study of the interplay between the two species in another publication.  For samples prepared as described above, we report the behavior of avalanches of the major species where the minor species plays no role, having already relaxed along the direction of the applied field.

Our studies of avalanches of the major species were carried out using the following experimental protocol.  After preparing the sample as described in the preceding paragraph, the magnetic field was ramped to a preassigned value in a direction opposite to the polarization of the major species, with the sample immersed in liquid $^3$He at $300$ mK.  It is important to note that this temperature is well below the blocking temperature of $\approx 3$ K so that below about $2$ T, there was negligible reduction of the magnetization by relaxation via tunneling as the field was swept through the resonant fields; the sample thus remained fully magnetized.  The wire heater was then turned on at fixed magnetic field, and the temperature of the sample was monitored  by measuring the resistance of the Ge thermometer using standard four-terminal techniques.

A typical curve showing the temperature as a function of time is shown in Fig. 2.  
\begin{figure}[htbp]
\begin{center}
  \includegraphics[width=3in]{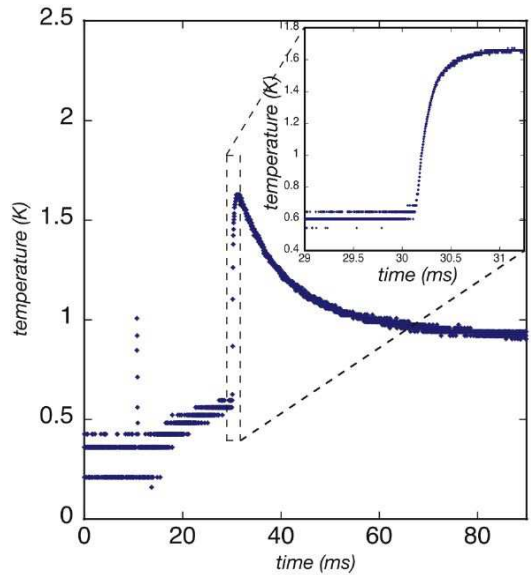}
\caption{Temperature recorded by the thermometer in contact with the crystal for an avalanche triggered at $1.85$ T.  The narrow peak at $13$ ms is electrical noise when the heater is turned on.  The abrupt rise at $30$ ms and $0.626$ K is due to heat released by the avalanche.  The inset shows data taken at the ignition temperature with higher resolution for the same avalanche \cite{foot}.}
\label{default}
\end{center}
\end{figure}
A spike occurs at $t \approx 13$ ms when the heater is turned on.  The subsequent slow rise in temperature between $t \approx 13$ and $t \approx 30$ ms reflects the gradual heating of the entire sample in response to the power provided by the heater.  The sharp rise in temperature at $t \approx 30$ ms signals the sudden release of heat associated with the ignition of an avalanche at a threshold temperature $T_{th}$  \cite{foot}.  The inset shows the data on an expanded time scale.  Measurements were repeated several times at a given field, and were reproducible within a given run.  Similar data were taken at many different (fixed) magnetic fields.

Figure 3 show the threshold temperature required to ignite avalanches plotted as a function of the magnetic field for fixed fields between $0.4$ T and $2.0$ T.   Sharp dips in the ignition temperature occur at magnetic fields denoted by vertical lines.  These magnetic fields correspond to thermally-assisted spin tunneling across the anisotropy barrier in Mn$_{12}$-ac  \cite{Friedman,abrupt}, effectively reducing the anisotropy barrier.  

\begin{figure}[htbp]
\begin{center}
  \includegraphics[width=3in]{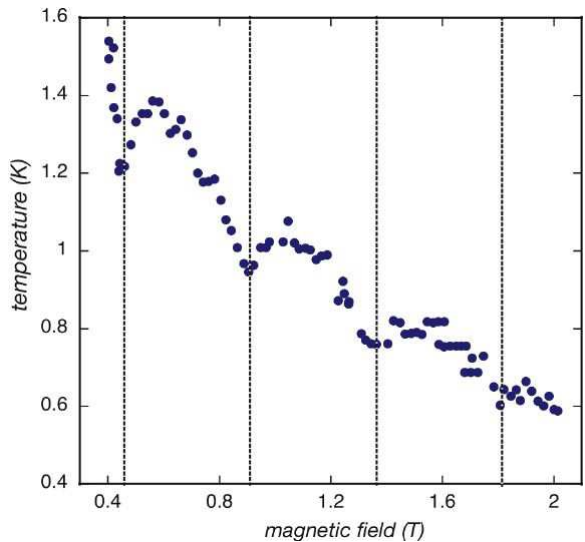}
\caption{Temperature required to ignite avalanches plotted as a function of magnetic field.  The vertical lines denote the magnetic fields where sharp minima occur in the ignition temperature corresponding to thermally-assisted tunneling near the top of the anisotropy barrier \cite{abrupt}.  The overall decrease in ignition temperature is due to the reduction of the anisotropy barrier as the field is increased. }
\label{default}
\end{center}
\end{figure}

\begin{figure}[htbp]
\begin{center}
  \includegraphics[width=3in]{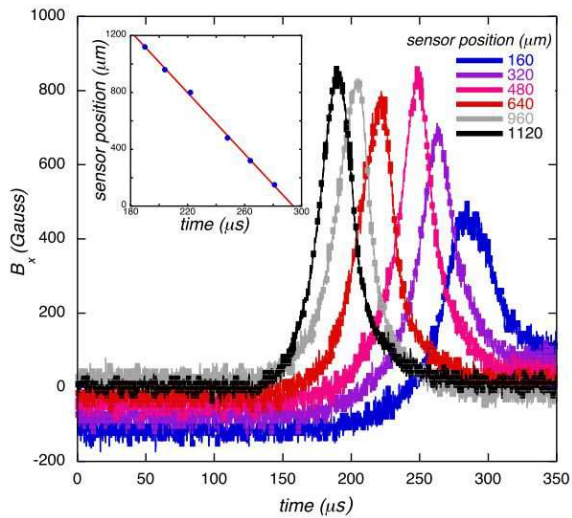}
\caption{Signals recorded by six Hall sensors in contact with the crystal for an avalanche triggered at $2$ T.  The inset shows sensor position versus the time at which the sensor recorded peak amplitude.  A straight line fit yields a velocity of $10.6$ m/s.}
\label{default}
\end{center}
\end{figure}

\begin{figure}[htbp]
\begin{center}
  \includegraphics[width=3in]{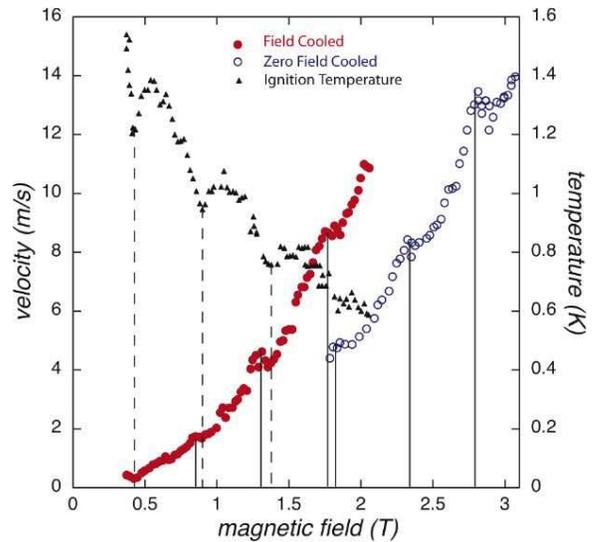}
\caption{Velocity of propagation of avalanches (right-hand y-axis) for field-cooled (filled circles) and zero-field-cooled (open circles) samples versus magnetic field at which avalanche was triggered.  The triangles show the ignition temperature (left-hand y-axis).  The solid vertical lines drawn from the bottom denote velocity maxima; the dashed vertical lines drawn from the top denote ignition temperature minima.  The overall increase of the velocity with increasing magnetic field is due to the decrease of the anisotropy barrier.}
\label{default}
\end{center}
\end{figure}

As mentioned earlier, the absolute value of the ignition temperature was reproducible within a given experimental run, but varied by as much as $0.25$ K from one run to another.  This is undoubtedly due to uncontrolled thermal gradients that were different depending on the thermal connection between the thermometer and the sample.  For example, the thickness of the layer of Apiezon M grease was perforce different for different runs.  It is important to note that strong minima were observed in $all$ runs at the same magnetic fields.  This behavior is robust and reproducible. 

Garanin and Chudnovsky have recently provided a detailed theoretical foundation for the newly-discovered process of magnetic deflagration by extending many of the results from the classical theory of combustion to the process of spin reversal in molecular magnets  \cite{Garanin}.  Consistent with the data shown in Fig. 3, their theory predicts a significant drop in the threshold temperature required to trigger avalanches at the "resonant" values of magnetic field where the barrier against spin reversal is effectively reduced due to resonant quantum spin tunneling.

We now present measurements of the magnetization obtained in separate experimental runs for similar Mn$_{12}$-ac crystals.  Time-resolved measurements of the local magnetization were obtained from measurements of the transverse component of the magnetic field, $B_x$, during an avalanche using six $30 \times 30 \mu$m$^2$ two-dimensional electron gas GaAs Hall sensors placed along the crystal to probe a significant fraction of its length.   Figure 4 shows the time of arrival at each sensor of the narrow interface between regions of the sample with anti-parallel magnetizations .  The velocity of propagation of the avalanche is then deduced from the known spacing between the sensors, as shown in the inset (see references \cite{Avraham, Suzuki} for experimental details).

The velocity of avalanches is shown in Fig. 5, where the filled circles denote results for avalanches that entail the reversal of the full magnetization from one direction to the other along the c-axis, and the open circles are for zero-field-cooled samples where the magnetization changes by half the amount, from zero to full magnetization.  The ignition temperatures shown in Fig. 3 are also plotted as triangles for comparison.  The vertical lines drawn in Fig. 5 denote the magnetic fields at which minima occur in the ignition temperature and maxima occur for the velocity.  It is interesting to note that the minimum ignition temperature occurs at higher field than the corresponding velocity maximum (see data at $H \approx 0.9$ T and $H \approx 1.35$ T).   In a similar manner, the velocity maximum for the zero-field-cooled sample at $H \approx 1.8$ T is at a slightly higher magnetic field than for the field-cooled case.  In both cases, the higher resonant field indicates that the tunneling takes place for energy level crossings that are deeper in the potential well \cite{abrupt}.  A detailed study of this effect will be published elsewhere.

The velocity of propagation increases as the field is raised and the barrier to spin reversal is reduced.  From measurements at $\approx 2.1$ K, Hernandez-Minguez {\it et al.}  \cite{Alberto2, Alberto3} have reported maxima at $0.9$ T and $1.35$ T, in agreement with predictions of the theory of Garanin and Chudnovsky \cite{Garanin}.  Although maxima are barely discernible at low magnetic fields at the lower (initial) temperatures of our experiments, they become evident at higher magnetic fields.   Further study is required to determine the conditions (e. g. temperature, size and direction of magnetic field) for observing the the effect of quantum tunneling on the velocity of propagation of avalanches.

To summarize: using fast local measurements of the temperature and time-resolved measurements of the local magnetization, we have shown that quantum tunneling of the magnetization plays a significant role in determining the threshold temperature for the ignition as well as the propagation of avalanches, as predicted by Garanin and Chudnovsky \cite{Garanin}.

We are grateful to Eugene Chudnovsky and Dmitry Garanin for many illuminating discussions.  This work was supported at City College by NSF grant DMR-00451605.  E. Z.. acknowledges the support of the Israel Ministry of Science, Culture and Sports.  Support for G. C. was provided by NSF grant CHE-0414555.


\begin{thebibliography}{5}

\bibitem{Lis}
{T. Lis}, Acta Cryst. {\bf B69}, 2042 (1980).

\bibitem{Sessoli}
{R. Sessoli, D. Gatteschi, A. Caneschi, and M. A. Novak}, Nature
(London) {\bf 365}, 141  (1993).

\bibitem{Friedman}
{J. R. Friedman, M. P. Sarachik, J. Tejada, and R. Ziolo}, Phys. Rev. Lett. {\bf 76},  3830  (1996); {J. R. Friedman, M. P. Sarachik, J. Tejada, J. Maciejewski, and R. Ziolo}, J. Appl. Phys.  {\bf 79}, 6031 (1996); {J. M. Hernandez, X. X. Zhang, F. Luis, J. Tejada, J. R. Friedman, M. P. Sarachik, and R. Ziolo}, Phys. Rev. B {\bf 55}, 5858 (1997).

\bibitem{Thomas}  L. Thomas, F. Lionti, R. Ballou, D. Gatteschi, R. Sessoli,
and B. Barbara, Nature (London) {\bf 383}, 145 (1996); J. M. Hernandez, X. X. Zhang, F. Luis, J. Bartolome, J. Tejada, and R. Ziolo, Europhys. Lett., {bf 35}, 301 (1996).

\bibitem{Barbara}
{B. Barbara, L. Thomas, F. Lionti, I. Chiorescu, and A. Sulpice}, J. Magn. Magn. Mater. {\bf 200}, 167 (1999).

\bibitem{Gatteschi}
{D. Gatteschi and R. Sessoli}, Angew. Chem., Int. Ed. {\bf 42}, 268 (2003)

\bibitem{Friedman2}
{J.R. Friedman}, in {\it Exploring the Quantum/Classical Frontier: Recent Advances in Macroscopic Quantum Phenomena}, edited by J.R. Friedman and S. Han (Nova Science, Hauppauge, NY, 2003), p. 179.

\bibitem{del Barco}
{E. del Barco, A.D. Kent, S. Hill, J.M. North, N.S. Dalal, E. Rumberger, D.N. Hendrikson, N. Chakov, and G. Christou}, J. Low Temp. Phys. {\bf 140}, 119 (2005).

\bibitem{Paulsen}
{C. Paulsen and J.G. Park}, in {\it Quantum Tunneling of Magnetization-QTM'94}, edited by L. Gunther and B. Barbara (Kluwer, Dordrecht, The Netherlands, 1995), pp. 189-207. 

\bibitem{thermal}
{F. Fominaya, J. Villain, P. Gandit, J. Chaussy, and A. Caneschi, Phys. Rev. Lett. {\bf 79}, 1126 (1997; J. Tejada, E.M. Chudnovsky, J.M. Hernandez, and R. Amigo}, Appl. Phys. Lett. {\bf 84}, 2373 (2004); A. Hernandez-Minguez, A. Jordi, R. Amigo, A. Garcia-Santiago, J.M. Hernandez, and J. Tejada, Europhys. Lett. {\bf 69}, 270 (2005); C.H. Webster, O. Kazakova, J.C. Gallop, P.W. Josephs-Franks, A. Hernandez-Minguez, A.Ya. Tzalenchuk, http://arxiv.org/abs/cond-mat/0609586.

\bibitem{Suzuki}
{Yoko Suzuki, M. P. Sarachik, E. M. Chudnovsky, S. McHugh, R. Gonzalez-Rubio, Nurit Avraham, Y. Myasoedov, E. Zeldov, H. Shtrikman, N.E. Chakov, and G. Christou }, Phys. Rev. Lett. {\bf 95}, 147201 (2005).

\bibitem{Glassman}  

{I. Glassman}, {\it Combustion} (Academic Press, New York, 1996).

\bibitem{Alberto2} 
{A. Hernandez-Minguez, J. M. Hernandez, F. Macia, A. Garcia-Santiago, J. Tejada, and P. V. Santos}, Phys. Rev. Lett. {\bf 95}, 217205 (2005).

\bibitem{Alberto3}
{A. Hernandez-Minguez, F. Macia, J. M. Hernandez, J. Tejada, and P. V. Santos}, http://arxiv.org/abs/cond-mat/0609429.

\bibitem{Wernsdorfer1}
{W. Wernsdorfer, R. Sessoli, D. Gatteschi}, Europhys. Lett. 47, 254 (1999). 

\bibitem{Wernsdorfer2}
{W. Wernsdorfer, N.E. Chakov, G. Christou}, http://arxiv.org/abs/cond-mat/0405014.

\bibitem{foot}
{The noise at low temperatures is exaggerated by two factors: a non-linearity in the thermometer that limits the resolution for temperatures below $0.4K$; and noise associated with digitizing (continuous) data acquired by the oscilloscopes.}

\bibitem{abrupt}

{A. D. Kent, Yicheng Zhong, L. Bokacheva, D. Ruiz, D. N. Hendrickson, and M. P. Sarachik, Europhys. Lett. {\bf 49}, 521 (2000); K. M. Mertes, Y. Zhong, M. P. Sarachik, Y. paltiel, H. Shtrikman, E.Zeldov, E. Rumberger, D. N. Hendrickson, and G. Christou, Europhys. Lett. {\bf 55}, 874 (2001).}

\bibitem{Garanin}
{D. A. Garanin and E. M. Chudnovsky, Phys. Rev. B {\bf 65}, 94423 (2002).}

\bibitem{Avraham}

{N. Avraham, A. Stern, Y. Suzuki, K. M. Mertes, M. P. Sarachik, E.
Zeldov, Y. Myasoedov, H. Shtrikman, E. M. Rumberger, D. N. Hendrickson,
N. E. Chakov, and G. Christou}, Phys. Rev. B. {\bf 72}, 144428 (2005).


\end{thebibliography}
\end{document}